\documentclass[sigconf]{acmart}

\usepackage{fancyhdr}
\usepackage{booktabs}
\usepackage{multirow}
\usepackage[normalem]{ulem}

\renewcommand\footnotetextcopyrightpermission[1]{}
\AtBeginDocument{%
  }

\setcopyright{acmlicensed}
\copyrightyear{2018}
\acmYear{2018}
\acmDOI{XXXXXXX.XXXXXXX}
\acmConference[]{}{}{}
\acmBooktitle{}
\AtBeginDocument{%
  \fancyhead[LE]{}%
  \fancyhead[RO]{}%
}
\acmISBN{978-1-4503-XXXX-X/2018/06}

\acmSubmissionID{6611}

\begin{document}

\title{SketchSong: Hierarchical Song Generation with Sketch Planning and Fine-Grained Multi-Track Modeling}

\author{Xiaoyue Duan\textsuperscript{*},
Nanxing Hu\textsuperscript{*\textdagger},
Yutang Feng\textsuperscript{*},
Xudong Yan,
Jiatao Chen,
Jinchao Zhang\textsuperscript{\textdaggerdbl},
Jie Zhou}
\affiliation{%
  \institution{Pattern Recognition Center, WeChat AI, Tencent Inc.}
  \country{}
}
\email{{lilmoonduan, neilnxhu, yutangfeng, jasonxdyan, wietchen, dayerzhang, withtomzhou}@tencent.com}

\renewcommand{\shortauthors}{Duan et al.}

\begin{abstract}
Recent song generation systems can synthesize realistic audio, yet generating complete songs remains challenging for two reasons. First, explicit song-level arrangement planning remains limited in existing methods, so models often need to organize overall arrangement development while generating low-level audio details. This often leads to incoherence in arrangements, such as weak section transitions and limited dynamic progression. Second, coarse modeling of different musical parts obscures their distinct roles and interactions, limiting arrangement richness of generated songs. In this paper, we present SketchSong, a hierarchical song generation framework that addresses these issues through song-level sketch planning and fine-grained multi-track modeling. Along the temporal dimension, SketchSong first predicts a compact sequence of high-level sketch tokens derived from compressed audio representations, and then generates audio tokens conditioned on these sketches. This coarse-to-fine process gives the model an explicit arrangement plan before detailed audio generation. Along the track dimension, SketchSong explicitly models four tracks, i.e., vocals, bass, drums and other instruments. This enables the model to capture the roles and interactions of different musical parts more precisely. Experiments on song generation benchmarks show that SketchSong consistently outperforms our baseline on both objective metrics and human listening tests. Despite not employing additional post-training for preference optimization such as lyrics and text-prompt alignments, SketchSong achieves competitive results against strong, post-trained open-source systems. Ablation studies further show that song-level sketch planning mainly improves long-range arrangement development and perceptual musical progression, while fine-grained multi-track modeling mainly enhances arrangement richness and production quality, demonstrating the effectiveness of our overall design.

\end{abstract}



\maketitle

\makeatletter
\renewcommand\@makefnmark{}%
\renewcommand\@makefntext[1]{\noindent #1}%
\renewcommand\thefootnote{}%
\footnotetext{\textsuperscript{*}These authors contributed equally.\\
\textsuperscript{\textdagger}Work done during an internship at WeChat AI, Tencent Inc.\\
\textsuperscript{\textdaggerdbl}Corresponding author.}
\makeatother

\begin{figure*}[t]
  \centering
  \includegraphics[width=\linewidth]{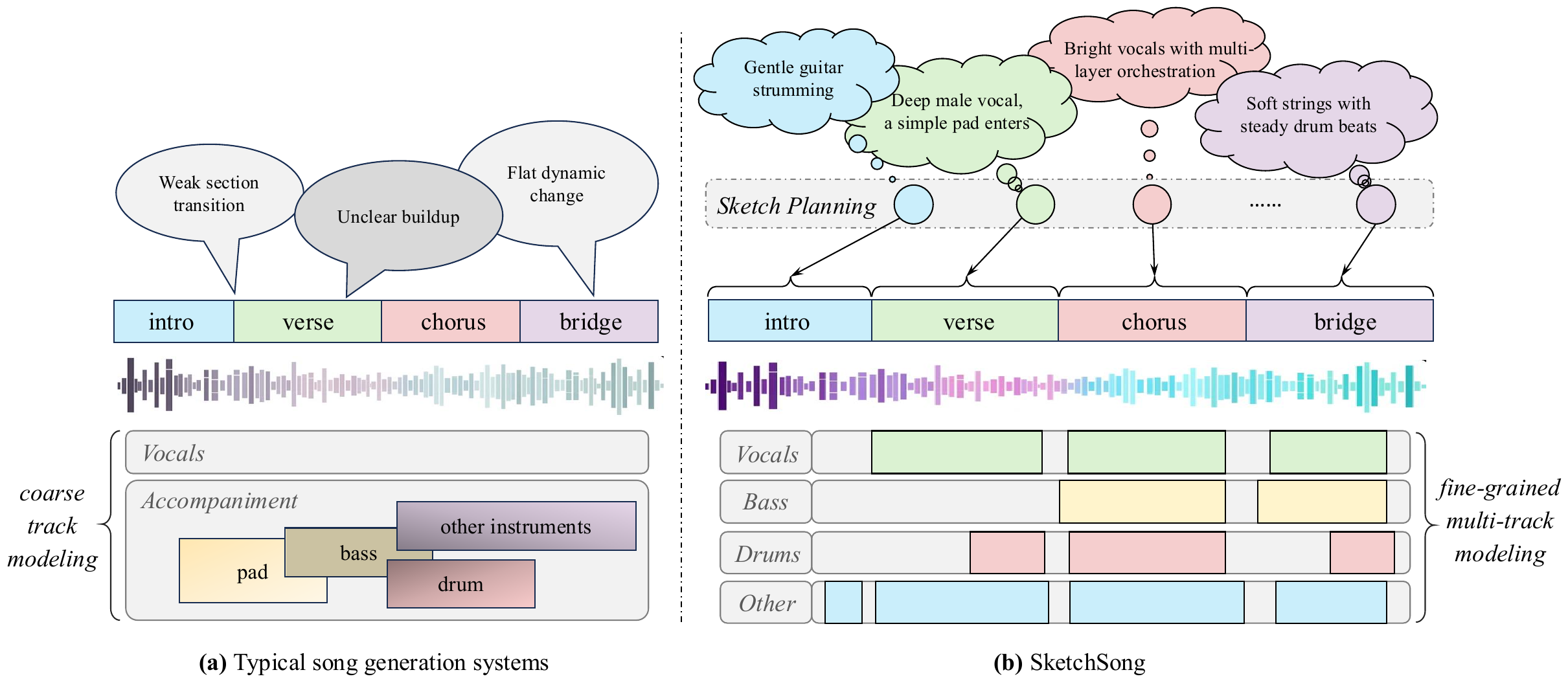}
  \caption{Limitations of existing song generation systems versus SketchSong. (a) Existing systems typically exhibit limited song-level planning and entangled instrumental roles inside accompaniment. (b) SketchSong introduces song-level sketch planning and fine-grained multi-track modeling, leading to clearer arrangement development and richer musical instrumentation.}
  \label{Fig.1}
\end{figure*}

\section{Introduction}

Recent advances in generative audio models have substantially improved conditional song synthesis from structured inputs such as lyrics, text descriptions, and reference audio~\cite{lei2024songcreator,ning2025diffrhythm,lei2025levo,yang2025songbloom}. However, generating a convincing complete song requires more than locally realistic sound. A good song needs clear section development, natural changes in energy, and coordinated interactions among vocals and instrumental parts. These qualities should be maintained throughout the song, rather than only within short local spans. Despite encouraging progress in recent song generation systems, achieving both global coherence and rich arrangement remains challenging.

One central challenge is limited song-level arrangement planning. Prior work has explored hierarchical semantic generation~\cite{agostinelli2023musiclm,prajwal2024musicflow}, CoT- or sketch-like intermediate guidance~\cite{lam2025analyzable,yang2025songbloom}, and segment-level prompting for structured control~\cite{cai2025segtune}. These directions suggest that intermediate abstractions can help, yet explicit planning for arrangement development in complete songs remains limited. As a result, models often need to handle both high-level arrangement development (\emph{e.g.}, section progression and dynamic contour) and low-level audio details within the same generation process. This can undermine coherence over the full song, leading to weak section transitions, limited dynamic progression, and less purposeful changes in instrumentation.

A second challenge lies in how different musical parts are modeled. Existing systems often generate songs as a single mixed signal~\cite{dhariwal2020jukebox,ning2025diffrhythm}, or adopt only coarse vocal--accompaniment decompositions~\cite{lei2024songcreator,lei2025levo}. Recent multi-track and stem-aware systems further suggest the value of explicit part modeling~\cite{yao2025jen,parker2024stemgen,zhang2025versatile}. In practice, different musical parts (\emph{e.g.}, vocals, bass, drums, and other instruments) do not contribute in the same way. Bass and drums strongly shape groove, rhythmic drive, and stylistic feel, while vocals and other instruments carry melody, harmony, and sonic texture. Therefore, better song generation benefits from finer part-aware modeling, so that the roles and interactions of different musical parts can be represented more precisely and used to support richer, better-coordinated arrangements.

To address these issues, we present SketchSong, a hierarchical song generation framework with song-level sketch planning and fine-grained multi-track modeling. For song-level planning, we compress high-level music representations into a low-rate sequence of discrete sketch tokens, and place them between the conditioning inputs and the audio token sequence. In this way, the model first predicts a sketch with high-level semantics before generating detailed audio, providing an explicit plan for arrangement development. We adopt a two-phase training strategy that first learns sketch prediction and then optimizes the full sketch-conditioned mixed-audio generation process. For track-aware modeling, we explicitly represent four tracks (\emph{i.e.}, vocals, bass, drums, and other instruments), enabling more precise modeling of musical roles and cross-part interactions. These two components are complementary: sketch planning improves how a song develops over time, while multi-track modeling improves how musical parts are organized within the arrangement.

We instantiate SketchSong in a two-stage song generation pipeline, following the strong mixed-to-track refinement paradigm used in a recent hierarchical song generation system~\cite{lei2025levo}. The first stage generates mixed audio tokens guided by song-level sketches, and the second stage refines the mixed audio tokens into finer-grained four-track tokens. We evaluate SketchSong on song generation benchmarks using both objective metrics and human listening tests. Experiments show that SketchSong consistently improves over our baseline and remains competitive with strong open-source systems, even without additional post-training for preference optimization or alignment. More importantly, the empirical results align well with our design motivations: song-level sketch planning mainly improves long-range arrangement development, section clarity, and perceived musical progression, while fine-grained multi-track modeling mainly enhances instrumental richness and production-related quality. Together, these results suggest that the two components are not redundant, but complementary in how they improve complete-song generation. In summary, the main contributions of this work are as follows.

\begin{itemize}
    \item We introduce song-level sketch planning for song generation by inserting compact discrete sketch tokens before audio generation, enabling a coarse-to-fine process that provides explicit guidance for arrangement development.
    \item We propose fine-grained multi-track modeling with four explicit tracks--vocals, bass, drums, and other instruments--to better capture the roles and interactions of different musical parts.
    \item We integrate these approaches into a hierarchical song generation framework. Experiments show that SketchSong consistently improves over our baseline on both objective metrics and human listening tests, while ablations further indicate that sketch planning benefits structure-related qualities and multi-track modeling benefits arrangement richness and production quality.
\end{itemize}

\section{Related Work}

\textbf{Song Generation.} Early song generation systems either modeled songs directly in raw or mixed audio, or relied on relatively coarse vocal--accompaniment decomposition. Jukebox~\cite{dhariwal2020jukebox} demonstrated hierarchical raw-audio song generation with VQ-VAE compression and autoregressive Transformers. More recent systems, including SongCreator~\cite{lei2024songcreator}, LeVo~\cite{lei2025levo}, DiffRhythm~\cite{ning2025diffrhythm}, DiffRhythm~2~\cite{jiang2025diffrhythm}, SongBloom~\cite{yang2025songbloom}, Muse~\cite{jiang2026muse}, and ACE-Step~\cite{gong2026ace}, have substantially improved full-song generation quality and controllability. Seed-Music~\cite{bai2024seed} presents a suite of hybrid autoregressive and diffusion systems supporting vocal music generation and interactive editing from multi-modal conditions. YuE~\cite{yuan2025yue} scales open foundation models for long-form music generation with large language model backbones. However, these systems generally do not combine explicit song-level planning with finer-grained track refinement in the way targeted by SketchSong.

\textbf{Hierarchical Planning and Structured Control.} Intermediate representations and structured controls have become an important direction in music generation. MusicLM~\cite{agostinelli2023musiclm} uses semantic-to-acoustic generation with MuLan-based conditioning. MusicFlow~\cite{prajwal2024musicflow} adopts a cascaded flow-matching design that first generates semantic features from text and then acoustic features from semantics, reinforcing the now-common pattern of splitting generation into a semantic stage and an acoustic stage. MusiCoT~\cite{lam2025analyzable} introduces CoT-style latent musical thoughts, where CLAP-derived latent representations serve as an analyzable intermediate planning sequence ahead of audio-token generation. SongBloom~\cite{yang2025songbloom} applies sketch-based intermediate generation to full songs, and SegTune~\cite{cai2025segtune} enables segment-level structured control. On the controllability side, JASCO~\cite{tal2024joint} combines global text control with local, time-aligned conditioning from symbolic or audio-derived signals such as chords, melody, and drum tracks. MusiConGen~\cite{lan2024musicongen} extends MusicGen~\cite{copet2023simple} with temporally aligned rhythm and chord control, while Mustango~\cite{melechovsky2024mustango} enriches textual prompts with music-theoretic attributes such as beats, tempo, key, and chord progression. These works show the value of intermediate abstractions for long-range coherence and controllability, but they either target general music generation, rely on externally specified controls, or do not connect song-level planning to later fine-grained track modeling. In contrast, SketchSong predicts compact song-level sketch tokens and uses them as an internal planning stage before detailed song synthesis.

\textbf{Part-Aware and Multi-Track Modeling.} Explicit modeling of musical parts is important for arrangement quality and controllability. SongCreator~\cite{lei2024songcreator} and LeVo~\cite{lei2025levo} separate vocals and accompaniment, with LeVo further coupling mixed-token and dual-track modeling in a two-stage framework built on a specialized music codec~\cite{xu2025mucodec}. Beyond dual-track formulations, JEN-1 Composer~\cite{yao2025jen} models conditional generation over multiple tracks for human--AI co-composition, StemGen~\cite{parker2024stemgen} generates missing stems conditioned on an existing context mix, and VersBand~\cite{zhang2025versatile} decomposes song creation into dedicated vocal and accompaniment modules. These works highlight the benefits of part-aware generation, but fine-grained track organization remains relatively limited in end-to-end song generation. SketchSong extends this direction by refining songs into four explicit tracks---vocals, bass, drums, and other instruments---while coupling this finer-grained track modeling with song-level sketch planning.

\begin{figure*}[t]
    \centering
    \includegraphics[width=0.9\linewidth]{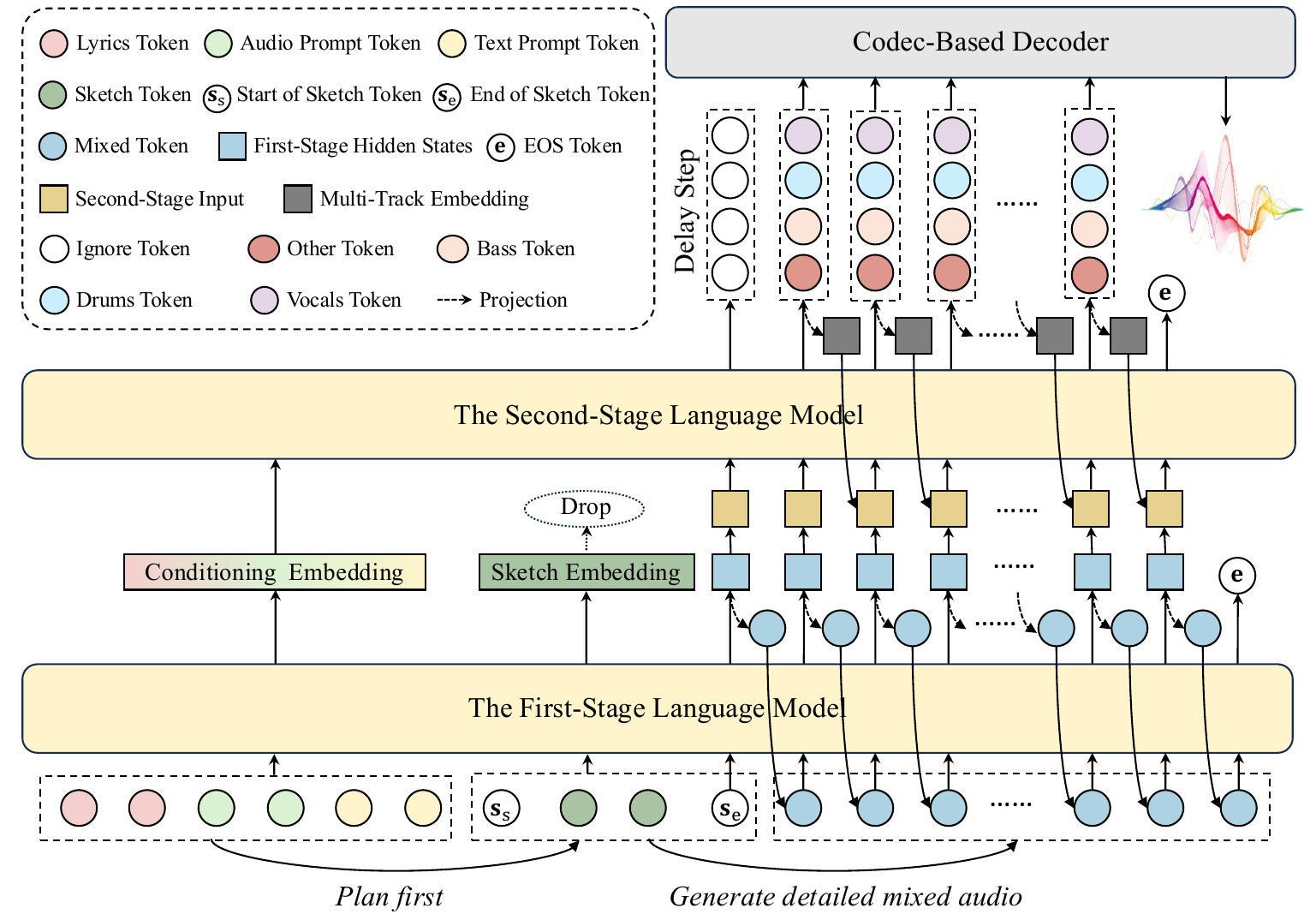}
    \caption{Overview of SketchSong. Given conditioning inputs, the first-stage language model predicts song-level sketch tokens before generating detailed mixed-song audio tokens. The second-stage language model then refines the mixed-song representation into four explicit tracks, \emph{i.e.}, vocals, bass, drums, and other instruments, with additional conditioning from first-stage hidden states which provide sketch-informed arrangement cues.}
    \label{fig:2}
\end{figure*}

\section{Method}

We present SketchSong, a hierarchical song generation framework for improving arrangement planning and track-aware refinement in full-song generation. SketchSong is built on the strong two-stage autoregressive generation paradigm of LeVo~\cite{lei2025levo}, where a first-stage language model generates an intermediate mixed-song representation, and a second-stage language model further refines it into separate vocal and accompaniment tracks. Within this general framework, SketchSong introduces two orthogonal methodological designs. First, we incorporate song-level sketch planning into the first generation stage. Second, we introduce a fine-grained four-track refinement stage over vocals, bass, drums, and other instruments. The overall framework is illustrated in Fig.~\ref{fig:2}.

\subsection{Overview}

Given structured lyrics, an optional text description, and an optional audio prompt, SketchSong generates a song in a hierarchical coarse-to-fine manner with two autoregressive language models, parameterized by $\theta_1$ and $\theta_2$, respectively. In the first stage, the model $f_{\theta_1}$ performs song-level planning and mixed-audio generation. Rather than directly predicting mixed audio tokens from the conditioning inputs, it first predicts a compact sequence of discrete sketch tokens and then generates mixed audio tokens conditioned on these sketches:
\begin{equation}
\mathbf{s} = f_{\theta_1}(\mathbf{c}), \qquad
\mathbf{x}_{\mathrm{mix}} = f_{\theta_1}(\mathbf{c}, \mathbf{s}),
\label{eq: first_stage}
\end{equation}
where $\mathbf{c}$ denotes the condition tensors, $\mathbf{s}$ denotes the song-level sketch tokens, and $\mathbf{x}_{\mathrm{mix}}$ denotes the mixed-song audio tokens. The sketch tokens are constructed offline by extracting MuQ-MuLan~\cite{huang2022mulan} audio features from 5-second non-overlapping windows and mapping these features to discrete code indices with residual vector quantization. During inference, the sketch tokens can either be generated by the model itself or provided by the user, enabling controllable generation.

In the second stage, the model $g_{\theta_2}$ refines the intermediate mixed representation into four explicit tracks. Specifically, it takes the same conditioning inputs together with the mixed-song representation from the first stage, and predicts four token streams corresponding to vocals, bass, drums, and other instruments:
\begin{equation}
\mathbf{x}_{\mathrm{trk}} = g_{\theta_2}(\mathbf{c}, \mathbf{x}_{\mathrm{mix}}),
\end{equation}
where $\mathbf{x}_{\mathrm{trk}}$ denotes the four-track tokens. This design enables finer part-aware generation and allows the model to represent the roles and interactions of different musical parts more explicitly.

After token prediction, we reuse the codec-based decoding pipeline of LeVo~\cite{lei2025levo} to reconstruct waveforms. Since dedicated codecs for bass, drums, and other tracks are not separately trained in the current system, these three tracks share the accompaniment codec during decoding. Our two designs are complementary: song-level sketch planning improves how the song develops over time, while four-track refinement improves how musical parts are organized within the arrangement.

\subsection{Song-Level Sketch Planning}

Direct autoregressive song generation requires the first-stage model to organize high-level arrangement development (\emph{e.g.}, section progression, dynamic contour, and layer evolution) while also predicting dense audio tokens within the same generation process, which makes long-range arrangement control difficult. To provide explicit planning guidance before detailed audio synthesis, we introduce song-level sketch planning, a low-rate discrete intermediate representation inserted between the conditioning inputs and the mixed-audio token sequence in the first-stage model.

\textbf{Song-Level Sketch Tokens.} We construct song-level sketch tokens offline from MuQ-MuLan~\cite{huang2022mulan} audio features. Specifically, for each training song, we extract a sequence of MuQ-MuLan feature vectors from non-overlapping 5-second windows, corresponding to a sketch frame rate of $0.2$ Hz. We choose MuQ-MuLan features because, as audio representations learned with audio-text contrastive supervision, they capture higher-level musical semantics than codec tokens while remaining temporally aligned with the evolving content of a song. This makes them suitable for representing coarse arrangement information such as section development, changes in density, and dynamic progression.

We then train a residual vector quantizer (RVQ) on these MuQ-MuLan audio features and use the resulting code indices as sketch tokens. In general, let each sketch frame be quantized by $L$ RVQ layers, so that the $t$-th sketch frame is represented as:
\begin{equation}
\mathbf{s}_t = \left[s_t^{(1)}, s_t^{(2)}, \ldots, s_t^{(L)}\right].
\end{equation}
The full song-level sketch sequence is written as $\mathbf{s} = \{\mathbf{s}_1, \mathbf{s}_2, \ldots, \mathbf{s}_T\}$.

\textbf{Sketch-Conditioned First-Stage Generation.} As illustrated in Eq.~\ref{eq: first_stage}, the first-stage model first predicts the sketch sequence from the conditioning inputs, and then generates mixed audio tokens conditioned on the predicted sketch. In our implementation, the sketch sequence is inserted after the condition representations and before the mixed-audio token sequence. For each sketch frame, we first apply a delay pattern following MusicGen~\cite{copet2023simple} to the code indices of the $L$ RVQ layers. We then convert the delayed indices of each layer into token embeddings, and sum the resulting layer embeddings to form the sketch representation for that frame. We further introduce learnable sketch-boundary embeddings, \emph{i.e.}, the start of sketch tokens $\langle \mathbf{s}_{\mathrm{s}} \rangle$ and the end of sketch tokens $\langle \mathbf{s}_{\mathrm{e}} \rangle$, so that the first-stage model processes the sequence in the following order:
\begin{equation}
\left[\mathbf{c};\; \langle \mathbf{s}_{\mathrm{s}} \rangle;\; \mathbf{s}^{\mathrm{emb}}_1, \ldots, \mathbf{s}^{\mathrm{emb}}_T;\; \langle \mathbf{s}_{\mathrm{e}} \rangle;\; \mathbf{x}_{\mathrm{mix}}\right],
\label{eq:lm1_input_sequence}
\end{equation}
where $\mathbf{s}^{\mathrm{emb}}_t$ denotes the summed embedding derived from the $L$ RVQ-layer indices in $\mathbf{s}_t$. This design encourages the model to establish a coarse song-level plan before generating detailed mixed audio tokens.

\textbf{Two-Phase Training.} As learning song-level planning and detailed mixed-audio generation simultaneously is difficult, we adopt a two-phase training strategy, \emph{i.e.}, the model first learns to predict a good sketch and then learns to generate audio conditioned on that sketch. The two phases described below both optimize the first-stage language model. In Phase 1, we optimize the model to predict the sketch sequence from the conditioning inputs alone. In other words, the model learns the mapping from the condition tensors (\emph{i.e.}, structured lyrics with an optional text description or audio prompt) to song-level sketch tokens before learning mixed-audio generation. In this phase, we use only the sketch-token cross-entropy loss $\mathcal{L}_{\mathrm{sk}}$ as supervision, \emph{i.e.}, $\mathcal{L}_{\mathrm{P1}}=\mathcal{L}_{\mathrm{sk}}$.

In Phase 2, we train the full sketch-conditioned mixed-audio generation process with joint sketch and audio supervision. That is, the model is trained on the full mapping from condition tensors to sketch tokens and then to mixed audio tokens. Let $\mathcal{L}_{\mathrm{mix}}$ denote the token-level cross-entropy loss over mixed audio tokens, and the objective of Phase 2 is:
\begin{equation}
\mathcal{L}_{\mathrm{P2}} = \mathcal{L}_{\mathrm{mix}} + \lambda_{\mathrm{sk}} \mathcal{L}_{\mathrm{sk}},
\label{eq:lm1_second_phase}
\end{equation}
where $\lambda_{\mathrm{sk}}$ denotes the weight of the sketch-token loss. This phase teaches the model to use the predicted sketch sequence as an intermediate planning scaffold for mixed-audio generation.

At inference time, SketchSong supports two modes. In the default mode, the first-stage model autoregressively predicts the song-level sketch sequence from the condition tensors, and then generates mixed audio tokens conditioned on that sketch. Alternatively, a user can provide a custom sketch sequence, or extract one from a reference song, which exposes song-level planning as an explicit control interface for arrangement-aware generation.

\subsection{Fine-Grained Multi-Track Modeling}

While song-level sketch planning improves how a song develops over time, full-song generation also requires finer modeling of how individual musical parts are organized in the arrangement. A mixed-song representation alone does not explicitly distinguish the roles of different musical parts within the accompaniment. We therefore model the second stage with four explicit tracks, \emph{i.e.}, vocals, bass, drums, and other instruments.

\textbf{Four-Track Target Streams.} Let $\mathbf{x}_{\mathrm{trk}}\!=\!\{\mathbf{x}_{\mathrm{voc}},\!\mathbf{x}_{\mathrm{bass}},\!\mathbf{x}_{\mathrm{drums}},\!\mathbf{x}_{\mathrm{other}}\}$ denote the four target token streams. Given the same condition tensors $\mathbf{c}$ as in the first stage, \emph{i.e.}, the representations derived from structured lyrics and an optional text description or audio prompt, the second stage takes the intermediate mixed-song tokens $\mathbf{x}_{\mathrm{mix}}$ as upstream context and predicts these four streams autoregressively. This finer decomposition is motivated by the fact that a single accompaniment stream entangles multiple instrumental layers, and explicitly separating important instruments (\emph{e.g.}, bass and drums) enables more explicit part-aware arrangement refinement.

\textbf{Sketch-Informed Hidden-State Conditioning.} To inject arrangement information from song-level sketch planning into the second stage, we condition the second-stage language model on the hidden states of the first-stage language model. Specifically, during second-stage training, we feed the sequence constructed in Eq.~\ref{eq:lm1_input_sequence} into the frozen first-stage language model using ground-truth sketch tokens $\mathbf{s}$ and ground-truth mixed-song tokens $\mathbf{x}_{\mathrm{mix}}$, and extract its last-layer hidden states. We denote the resulting hidden-state sequence as:
\begin{equation}
\mathbf{H}_1=f_{\theta_1}^{\mathrm{hid}}(\mathbf{c}, \mathbf{s}, \mathbf{x}_{\mathrm{mix}})=\left[\mathbf{H}_{\mathrm{c}};\; \mathbf{H}_{\mathrm{s}};\; \mathbf{H}_{\mathrm{a}}\right],
\end{equation}
where $\mathbf{H}_{\mathrm{c}}$, $\mathbf{H}_{\mathrm{s}}$, and $\mathbf{H}_{\mathrm{a}}$ denote the hidden states corresponding to the condition, sketch, and mixed-audio segments, with their sequence lengths denoted as $T_c$, $T_s$, and $T_a$, respectively. Since the input of the second-stage language model does not contain sketch tokens, we discard the sketch segment and use the aligned hidden-state sequence $\tilde{\mathbf{H}}_1=[\mathbf{H}_{\mathrm{c}};\; \mathbf{H}_{\mathrm{a}}]$, with a length of $T_c+T_a$.

\textbf{Multi-Track Input and Prediction.} For the second-stage language model, the token streams of vocals, bass, drums, and other instruments are first embedded by four track-specific embedding layers, respectively, and then summed to form a multi-track representation. This multi-track representation is concatenated after the condition sequence to form an intermediate second-stage sequence. We then concatenate this intermediate sequence with $\tilde{\mathbf{H}}_1$ along the feature dimension and map the result back to the hidden size expected by the second-stage model through an MLP, yielding the final input sequence of the second-stage model. In this way, the second-stage language model is jointly conditioned on the condition tensors, the multi-track token context, and the first-stage hidden representation of sketch-conditioned mixed-song generation.

The output hidden states of the second-stage language model are denoted by $\mathbf{H}_2$. We apply four track-specific linear heads to $\mathbf{H}_2$ to predict the token distributions of vocals, bass, drums, and other instruments. Let $\mathcal{L}_{\mathrm{voc}}$, $\mathcal{L}_{\mathrm{bass}}$, $\mathcal{L}_{\mathrm{drums}}$, and $\mathcal{L}_{\mathrm{other}}$ denote the corresponding token-level cross-entropy losses, and the training objective is computed as:
\begin{equation}
\mathcal{L}_{\mathrm{trk}} = \frac{1}{4}\left(\mathcal{L}_{\mathrm{voc}} + \mathcal{L}_{\mathrm{bass}} + \mathcal{L}_{\mathrm{drums}} + \mathcal{L}_{\mathrm{other}}\right).
\label{eq:second_stage_loss}
\end{equation}
Thus, the second stage is supervised only by the four-track prediction loss, while the intermediate mixed-song representations provide coarse musical context and sketch-informed arrangement cues through the hidden-state conditioning from the first-stage language model.

\subsection{Training and Inference}

SketchSong is trained in a stage-wise manner. We first optimize the first-stage language model with the aforementioned two-phase training strategy. In Phase 1, the model predicts sketch tokens from the conditioning inputs, and is supervised only by $\mathcal{L}_{\mathrm{P1}}$. We then continue training the same model in Phase 2 on the full sketch-conditioned mixed-audio generation process, where the model is supervised by $\mathcal{L}_{\mathrm{P2}}$ in Eq.~\ref{eq:lm1_second_phase}. After the first-stage model is trained, we then train the second-stage language model for four-track prediction with the loss defined in Eq.~\ref{eq:second_stage_loss} while keeping the first-stage model frozen for hidden-state conditioning.

At inference time, SketchSong supports two modes at the first stage. In the default mode, the first-stage model predicts a song-level sketch sequence from the condition tensors, and then generates mixed-song audio tokens conditioned on that sketch. Alternatively, a user can provide a sketch sequence, in which case the first-stage model directly generates mixed-song audio tokens conditioned on the user-specified sketch. After the first stage finishes, the second-stage language model takes the same conditioning inputs together with the mixed-song representation generated by the first stage, and autoregressively predicts four-track tokens for vocals, bass, drums, and other instruments.

Finally, we reuse the same MuCodec-based waveform reconstruction pipeline as LeVo~\cite{lei2025levo} to decode the predicted multi-track tokens into audio. The vocal stream is decoded with the vocal codec, whereas the bass, drums, and other streams are decoded with the accompaniment codec, since dedicated codecs for these three tracks are not separately trained in the current system.

\begin{table*}[t]
\centering
\caption{Comparison of different methods on objective evaluation metrics. CE/CU/PC/PQ denote content enjoyment, content usefulness, production complexity, and production quality; Coh./Mus./Mem./Cla./Nat. denote coherence, musicality, memorability, clarity, and naturalness. We include both the released LeVo checkpoint and a LeVo model trained on our dataset and setup, as the latter serves as our direct baseline. Note that all open-source methods (except YuE) adopt post-training, while our current SketchSong does not integrate post-training. Best and second-best results in each column are marked in bold and underline, respectively.}
\vspace{-2mm}
\label{tab:main_results}
\small
\setlength{\tabcolsep}{4pt}
\renewcommand{\arraystretch}{1.2}
\begin{tabular}{l c c c c c c c c c c c c}
\toprule
\multirow{2}{*}{Methods} 
& \multirow{2}{*}{FAD $\downarrow$} 
& \multirow{2}{*}{MuQ-T $\uparrow$} 
& \multirow{2}{*}{PER $\downarrow$}
& \multicolumn{4}{c}{AudioBox Aesthetics $\uparrow$}
& \multicolumn{5}{c}{SongEval Aesthetics $\uparrow$} \\
\cmidrule(lr){5-8} \cmidrule(lr){9-13}
& & & 
& CE & CU & PC & PQ 
& Coh. & Mus. & Mem. & Cla. & Nat. \\
\midrule
YuE                  &   4.16   &  0.17    &   45.16   &  7.02    &  7.46    &   4.86   &   7.85   &    3.34  &    3.16  &    3.21  &   3.16   &   3.08   \\
DiffRhythm 2          &   3.49   &  \textbf{0.40}    &  25.6    &   7.51   &  7.78    &  5.68    &   8.19   &   \underline{3.95}   &   3.67   &   3.79   &    \underline{3.68}  &    3.54  \\
ACE-Step 1.5         &   \textbf{2.62}   &  0.27    &   \underline{24.5}   &  7.57    &  7.88    &  5.86    &   8.21   &  \textbf{4.16}    & \textbf{4.11}     &  \textbf{4.09}    &  \textbf{4.04}    &  \textbf{3.73}    \\
LeVo (released)   &   3.79   &  0.24    &  \textbf{9.3}    &  7.71    &  7.93    &  5.85    &  8.40    &  3.92    &  \underline{3.75}    &  \underline{3.83}    &  3.66    &   \underline{3.58}   \\
\midrule
LeVo (our trained)  &   3.73   &  0.29    &   32.3   & 7.61     &   7.83   &   5.80   &   8.35   &  3.52    &    3.42  &   3.38   &  3.36    &    3.25  \\
+ Sketch planning      &  \underline{3.05}   &   \underline{0.33}   &  29.7    &   \underline{7.72}   &  \textbf{7.95}    &   \underline{6.11}   &  8.31    &  3.49    &    3.69  &   3.65   & 3.60     & 3.51     \\
+ Multi-track modeling&    3.82   &   0.29   &   31.4   & 7.58     &  7.90    &   5.98   &  \textbf{8.43}    &   3.88   &   3.35   &  3.36    &   3.28   &   3.20   \\
\textbf{SketchSong}            &   3.06   &   0.32   &   27.8   &   \textbf{7.76}   &   \underline{7.94}   &   \textbf{6.27}   &  \underline{8.42}    &   3.84   &  3.59    &  3.67    &  3.65    &   3.48   \\
\bottomrule
\end{tabular}
\end{table*}

\begin{table}[t]
\centering
\caption{MOS listening test results on subjective metrics. OVL/SSC/AD/IR/VQ denote overall quality, song structure clarity, arrangement development, instrumental richness, and vocal quality, respectively. Best and second-best results in each column are marked in bold and underline, respectively.}
\vspace{-3mm}
\label{tab:ablation_mos}
\small
\setlength{\tabcolsep}{8pt}
\renewcommand{\arraystretch}{1.25}
\begin{tabular}{lccccc}
\toprule
\multirow{2}{*}{Method} & \multicolumn{5}{c}{MOS $\uparrow$} \\
\cmidrule(lr){2-6}
 & OVL & SSC & AD & IR & VQ \\
\midrule
LeVo (our trained)   &   3.60   &   3.76   &  3.62    &  3.56    &  \underline{3.82}    \\
+ Sketch planning      &   \underline{3.64}   &   \textbf{3.88}   &  \textbf{3.90}    &  3.62    &  \textbf{3.88}    \\
+ Multi-track modeling&   3.58   &   3.70   &  3.60    &  \underline{3.78}    &  3.78    \\
SketchSong            &   \textbf{3.68}   &   \underline{3.84}   &  \underline{3.87}    &  \textbf{3.84}    &  3.80    \\
\bottomrule
\end{tabular}
\end{table}

\section{Experiments}

\subsection{Experimental Setup}

\textbf{Dataset.} We train SketchSong on a corpus of 1 million songs (approximately 54k hours). Following the automated SongPrep preprocessing pipeline~\cite{tan2025songprep}, we construct structured lyrics, section-level structure annotations, and separated tracks for each song. Specifically, we use Demucs~\cite{rouard2023hybrid} to decompose each song into four tracks, \emph{i.e.}, vocals, bass, drums, and other instruments. For music structure annotation, we adopt an All-In-One model enhanced with Dual-Path RNN (DPRNN) blocks~\cite{kim2023all} to predict section boundaries, section labels, and their timestamps. For lyric transcription, we directly use the SongPrep-provided recognition pipeline, which combines WER-FIX-based lyric refinement, a Zipformer ASR model~\cite{yao2023zipformer} with a dual-head ASR framework. We further apply a wav2vec 2.0-based word alignment module~\cite{baevski2020wav2vec} to calibrate structural outputs and correct timestamp mismatches. In addition, we generate a open-vocabulary text description for each song using Qwen2.5-Omni~\cite{xu2025qwen25omnitechnicalreport}.

\textbf{Implementation Details.} Our implementation follows the two-stage autoregressive framework of LeVo~\cite{lei2025levo}, while introducing song-level sketch planning in the first stage and four-track refinement in the second stage. We train both language models from scratch. For sketch token construction, we extract MuQ-MuLan audio features from non-overlapping 5-second windows, and train an RVQ codebook over these features. The RVQ adopts four quantizers (\emph{i.e.}, $L=4$), a codebook size of 4096, and an embedding dimension of 512. Its encoder and decoder both contain four ResMLP layers~\cite{touvron2022resmlp} with a hidden dimension of 2048. We train the RVQ for 50 epochs with batch size 512 and learning rate $3\times10^{-4}$, using cosine reconstruction loss together with commitment loss. The learned code indices are then used as song-level sketch tokens. The two language models are both trained on 64 NVIDIA A100 GPUs with batch size 2 per GPU. For the first-stage language model, both Phase 1 and Phase 2 are trained for 100k steps. The second-stage language model is trained for 50k steps. All stages use 4k warmup steps. The embedding layers for conditioning inputs, sketch tokens, and mixed-audio or four-track tokens are all trainable. Components not directly related to our two main contributions, including the codec-based waveform decoder pipeline, are kept fixed from the open-source checkpoint of LeVo. Unless otherwise stated, other settings follow LeVo, and more detailed configurations are provided in the supplementary material.

\textbf{Evaluation.} We evaluate SketchSong with both objective and subjective metrics. For objective evaluation, we report Fr\'echet Audio Distance (FAD)~\cite{kilgour2018fr} and Phoneme Error Rate (PER). To compute PER, we first extract the vocal track with Demucs~\cite{rouard2023hybrid} and then transcribe it using Whisper-large-v2~\cite{radford2023robust}. We also measure the similarity between the generated song and the input text description (denoted as ``MuQ-T'') based on pretrained MuQ-MuLan representations~\cite{huang2022mulan}. For aesthetic evaluation, we report: 1) Audiobox-Aesthetic~\cite{tjandra2025meta}, including content enjoyment (CE), content usefulness (CU), production complexity (PC), and production quality (PQ); and 2) SongEval~\cite{yao2025songeval}, including coherence, musicality, memorability, clarity, and naturalness. For these objective metrics, we randomly sample 100 Chinese songs and 100 English songs from the validation set to provide conditioning inputs for generating songs. For subjective evaluation, we conduct a mean opinion score (MOS) listening test. Each listener is randomly assigned 10 test cases, and samples from different methods for the same case share identical conditioning inputs. We invite 5 music experts and 10 amateur listeners to rate generated songs on a 1--5 scale over five dimensions, \emph{i.e.}, overall quality (OVL, focusing on musicality and naturalness), song structure clarity (SSC), arrangement development (AD), instrumental richness (IR), and vocal quality (VQ).

\textbf{Comparison Systems.} Our primary baseline is LeVo~\cite{lei2025levo}. Since the original LeVo training data are not publicly available, and LeVo also includes an additional DPO-based post-training stage that is not used in the current SketchSong system, we re-train a LeVo baseline on our own dataset under the same setup for fair comparison. We also include the performance of the released LeVo checkpoint in the experimental results. Beyond LeVo, we compare with several open-source song generation systems, including YuE~\cite{yuan2025yue}, DiffRhythm~2~\cite{jiang2025diffrhythm}, and ACE-Step~1.5~\cite{gong2026ace}. For these open-source baselines, we use their released checkpoints and default configurations.

\vspace{-1.5mm}
\subsection{Objective Results}

\textbf{Main Results.} Table~\ref{tab:main_results} jointly reports results against open-source comparison systems and our internal ablations. Compared to LeVo trained under our setup, SketchSong improves on most metrics, reducing FAD from 3.73 to 3.06 and PER from 32.3 to 27.8, while improving MuQ-T from 0.29 to 0.32. It also achieves the best aesthetic scores on CE and PC, and the second-best scores on CU and PQ, among all compared methods. Notably, all open-source comparison systems (except YuE) in Table~\ref{tab:main_results} are evaluated with their released post-trained checkpoints, whereas the current SketchSong does not use additional post-training such as dedicated lyric alignment, text-prompt alignment, or reward-based aesthetic optimization. As a result, while still lagging behind some released systems on PER and several SongEval dimensions, SketchSong is already competitive on FAD and several aesthetic metrics, demonstrating the effectiveness of our design. More comparisons are included in the supplementary.

\textbf{Ablation Study.} The ablation rows further show that the two proposed components contribute in different ways. Adding sketch planning on top of LeVo leads to the largest gains on FAD, CE, CU, musicality, and memorability, which is consistent with our motivation that song-level sketch planning mainly improves long-range arrangement development, thus generating songs with clearer progression and stronger overall musicality and memorability. By contrast, introducing fine-grained multi-track modeling on top of LeVo's dual-track paradigm most clearly improves production complexity and production quality, matching our expectation that finer part decomposition mainly benefits arrangement richness and production realism. When combined, the full SketchSong model achieves the best overall trade-off across these two dimensions.

At the same time, the ``+ Multi-track modeling'' variant alone does not improve FAD and only yields limited gain on PER. We attribute this to two reasons. First, multi-track modeling mainly refines the accompaniment, so its direct effect on vocal transcription accuracy is limited. Second, in the current system only the second-stage language model is made four-track-aware, while the generated bass, drums, and other tracks are still decoded by the shared LeVo accompaniment codec and then remixed with a simple $1/3$ amplitude scaling for each stream. This likely leads to less natural mixtures after remixing and can hurt FAD. Training a dedicated multi-track codec and diffusion decoder for separated tracks is a promising direction for improving overall generation quality, which we point to as future work.

\subsection{Subjective Results}

Table~\ref{tab:ablation_mos} reports the MOS listening test results. Compared with LeVo trained under our setup, SketchSong improves the overall quality score from 3.60 to 3.68 and the instrumental richness score from 3.56 to 3.84, indicating that the proposed design yields a clear perceptual benefit in both overall listening experience and arrangement richness. It also remains competitive on song structure clarity, arrangement development, and vocal quality.

The subjective ablations again reveal complementary effects of the two designs. Adding sketch planning yields the clearest gains on song structure clarity, arrangement development, and vocal quality, which aligns with the role of sketch planning in clarifying section progression and improving perceived song development. Adding multi-track modeling most clearly improves instrumental richness, suggesting that explicit bass, drums, and other-instrument modeling helps listeners perceive denser and more structured accompaniment without reducing much vocal quality. The full SketchSong model combines these advantages, achieving the best overall quality and instrumental richness while remaining close to the best sketch-only variant on structure-related metrics.

\begin{figure}[t]
    \centering
    \includegraphics[width=0.9\linewidth]{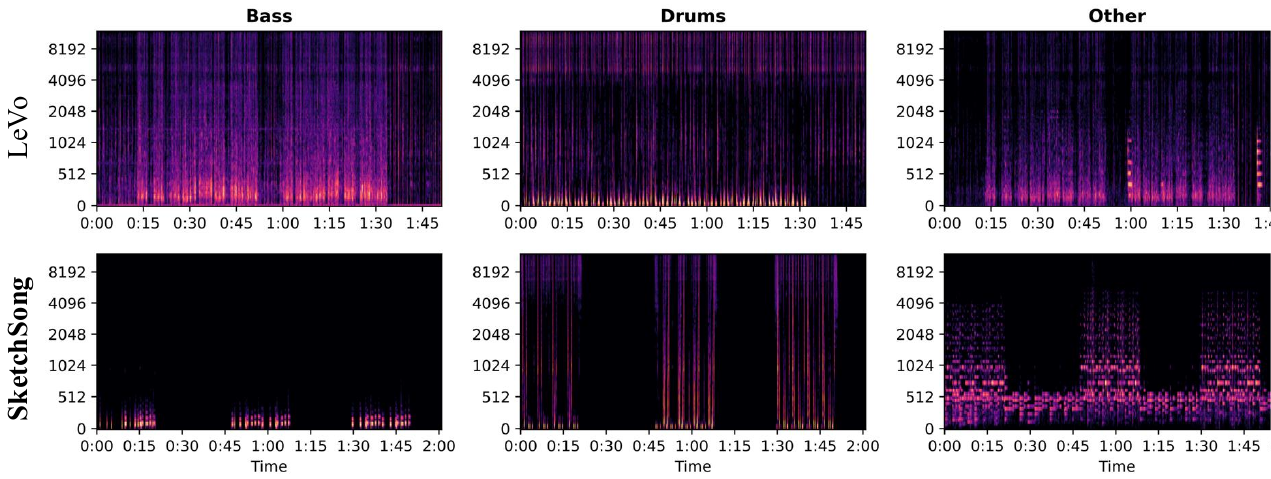}
    \vspace{-1.5mm}
    \caption{Case study on separated mel-spectrograms of generated songs. For the same test case, we first generate a song with LeVo and SketchSong, respectively, and then re-apply source separation to each generated waveform to obtain the bass, drums, and other-instrument tracks for visualization. Compared with LeVo, SketchSong produces clearer frequency-role separation across tracks and more structured time-varying arrangement development over the song.}
    \label{fig:multi-track-case-study}
\end{figure}

\begin{figure*}[t]
    \centering
    \includegraphics[width=0.8\linewidth]{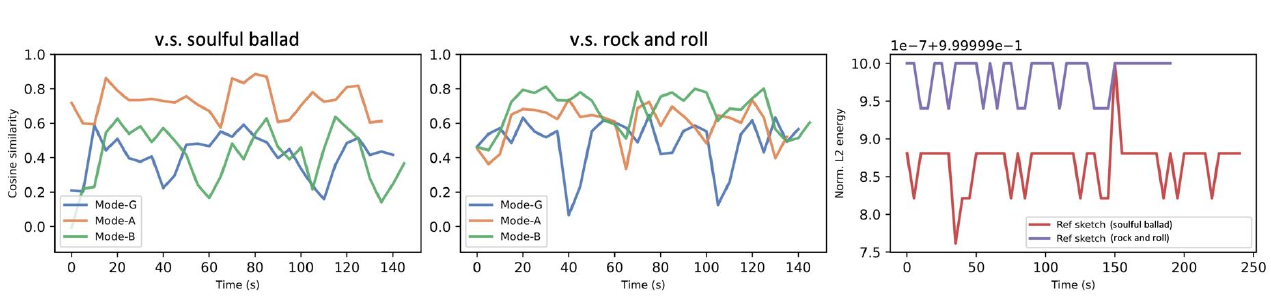}
    \caption{Sketch controllability under three inference modes. With the same text prompt (\emph{i.e.}, \emph{hip-hop}) fixed, replacing the sketch with a \emph{soulful ballad} or \emph{rock \& roll} reference shifts the generated audio toward the corresponding reference in MuQ-MuLan similarity and energy, showing that sketch substitution provides effective control over the final song.}
    \label{fig:sketch_control}
\end{figure*}

\subsection{Case Studies on Generated Audio and Sketch Controllability}

\textbf{Mel-Spectrogram Visualization.} Fig.~\ref{fig:multi-track-case-study} provides a qualitative case study to further illustrate how SketchSong affects the generated audio. 

From the perspective of fine-grained multi-track modeling, the contrast is clear. In the LeVo result, the separated bass spectrogram spreads across almost the full frequency range, which is not physically plausible for a bass-dominant track, since bass content should mainly concentrate in low frequencies. This suggests that bass-related content is entangled with other instrumental components and is therefore difficult to separate cleanly, matching the coarse-track motivation illustrated in Fig.~\ref{Fig.1}(a). By contrast, in SketchSong the separated bass, drums, and other-instrument tracks are much more clearly organized: the bass is mainly concentrated in low frequencies, while the drums cover a broader frequency range, which is reasonable because drums include both low-frequency kick components and high-frequency percussive content such as hi-hats and cymbals. This indicates that fine-grained multi-track modeling helps the model generate more explicit and separable instrumental roles, especially for bass and drums.

The same case also reveals the effect of song-level sketch planning on arrangement development over time. In the LeVo result, the spectrograms of different instrumental tracks remain relatively flat across the full duration, with little obvious structural change along the time axis. In other words, the instrumental energy tends to remain spread over the whole song without clear section-dependent variation. In SketchSong, however, the time-varying structure is much more apparent. For example, in the other-instrument track, the intro section (roughly 0:00--0:20) and the chorus sections (around 0:50--1:10) show fuller and richer spectral coverage, reflecting denser instrumentation and stronger emotional buildup; these sections are also accompanied by more active bass and drums. By contrast, in the verse sections (roughly 0:20--0:50 and 1:10--1:30), the spectral energy is noticeably sparser and mainly concentrated in the mid-to-low range, which is consistent with a softer pad-like accompaniment and the near absence of bass and drums. This suggests that sketch planning helps the model organize section-level development more purposefully, leading to clearer changes in density, instrumentation, and emotional progression across the song.

\textbf{Sketch Control.} To further verify that song-level sketch planning provides user-level controllability, we compare three inference modes while keeping the same text prompt fixed: \emph{hip-hop}. In \textbf{Mode-G}, the model predicts its own sketch from the text prompt and then generates songs conditioned on that sketch. In \textbf{Mode-A}, the text prompt remains unchanged, but the sketch is replaced with a fixed sketch extracted from a \emph{soulful ballad} song. In \textbf{Mode-B}, the text prompt also remains unchanged, but the sketch is replaced with a fixed sketch extracted from a \emph{rock \& roll} song. Fig.~\ref{fig:sketch_control} visualizes how these different sketch conditions influence the final generated audio through MuQ-MuLan-based analyses.

The left and middle panels show reference-audio similarity under the three modes. In the left panel, we compute the MuQ-MuLan audio-feature similarity between each generated song and the original \emph{soulful ballad} reference song. Mode-A consistently produces the highest similarity, even though the text prompt itself does not describe a soulful ballad. In the middle panel, we instead compute the similarity to the original \emph{rock \& roll} reference song, and Mode-B becomes the highest. This indicates that replacing the sketch alone can strongly steer the style of the generated song toward the reference associated with that sketch, even when the text prompt remains unchanged and does not explicitly match the injected sketch style.

The right panel further shows that sketch replacement also changes the energy profile of the generated audio. Comparing the MuQ-MuLan feature energy of the generated songs conditioned on the \emph{soulful ballad} sketch and the \emph{rock \& roll} sketch, the rock-conditioned results exhibit substantially higher energy than the soulful-ballad-conditioned results. This is consistent with the expected difference between the two sketch styles and further supports that the sketch does not merely affect a subtle latent representation, but has a strong and measurable impact on the final generated audio. Taken together, these results provide direct evidence that song-level sketch planning functions as an effective control interface: users can modify the overall style and high-level musical development of the generated song by replacing the sketch, even under the same text prompt.

\section{Conclusion}

In this paper, we presented SketchSong, a hierarchical song generation framework designed to improve full-song generation from two complementary aspects: long-range arrangement development and fine-grained part-aware modeling. To address the difficulty of organizing high-level song development while generating dense audio tokens, SketchSong introduces song-level sketch planning into the first stage, so that the model first predicts compact sketch tokens and then generates mixed-song audio conditioned on this plan. To better model musical roles and interactions within an arrangement, SketchSong further extends the second stage from a coarse vocal--accompaniment formulation to fine-grained four-track modeling over vocals, bass, drums, and other instruments. Experiments show that SketchSong consistently improves over a strong LeVo baseline trained under the same setup, and achieves competitive performance against open-source systems on both objective metrics and human listening tests. The ablation results further confirm that sketch planning mainly improves long-range coherence and arrangement development, while fine-grained multi-track modeling mainly enhances arrangement richness and production quality. The current system still has limitations, including the lack of additional post-training and the reuse of the accompaniment-side decoding pipeline for multiple non-vocal tracks. Future work can explore dedicated post-training, multi-track codec learning, and stronger waveform decoding for further quality and controllability improvements.

\bibliographystyle{ACM-Reference-Format}
\bibliography{sample-base}

\end{document}